\documentclass[fleqn,10pt]{SelfArx} % Document font size and equations flushed left

\usepackage{lipsum} % Required to insert dummy text. To be removed otherwise

\definecolor{color1}{RGB}{0,0,90} % Color of the article title and sections
\definecolor{color2}{RGB}{0,20,20} % Color of the boxes behind the abstract and headings

\usepackage{hyperref} % Required for hyperlinks
\hypersetup{colorlinks=true,urlcolor=blue,linkcolor=blue,linkbordercolor=blue,breaklinks=true,bookmarksopen=false,pdftitle={Title},pdfauthor={Author}}
\JournalInfo{Geomagnetism and Aeronomy, 2020, Vol. 60, No. 7} % Journal information
\Archive{}

\PaperTitle{
LONG-TERM ACTIVITY IN PHOTOSPHERES OF LOW-MASS STARS 
WITH STRONG MAGNETIC FIELDS
} % Article title

\Authors{N.\,I.~Bondar'\textsuperscript{1*}, M.\,M.~Katsova\textsuperscript{2**}
} % Authors

\affiliation{\textsuperscript{1}\textit{
Crimean Astrophysical Observatory RAS, Nauchny, Russia
}} % Author affiliation
\affiliation{\textsuperscript{2}\textit{
Sternberg State Astronomical Institute, Lomonosov Moscow State University, Moscow, Russia
}} % Author affiliation
\affiliation{\textsuperscript{*}e-mail: otbn@mail.ru}
\affiliation{\textsuperscript{**}e-mail: maria@sai.msu.ru}
\affiliation{Received: March 07, 2020. Accepted: April 29, 2020.}

\Keywords{stars: activity -- stars: cycles -- stars: starspots -- techniques: photometry}
\newcommand{\keywordname}{Keywords} 
\Abstract{
The behavior of the average annual luminosity of K--M dwarfs OU Gem, EQ Vir, 
V1005 Ori and AU Mic was studied at time intervals of several decades. 
The main sources of photometric data for 1989--2019 were the Hipparcos, 
ASAS, KWS databases. An analysis of long-term series showed that the 
average annual brightness of all stars varies cyclical. The durations of 
possible cycles for OU Gem, V1005 Ori and AU Mic are 40--42 years, for 
EQ Vir -- 16.6 years, and amplitudes of cycles are of 0.09--0.2$^m$. 
The selected stars belong to the group of red dwarfs for which the 
average surface magnetic field $\langle B\rangle$ exceeds of several 
kilogauss. We examined the type of the relationship between the 
parameters of the cycle, its duration and amplitude, for 9 stars with 
$\langle B\rangle < 4\;$kG. A tendency to increasing of the amplitude and duration 
of the cycle when the value of $\langle B\rangle$ decrease has been noted. It may be 
suggested that with an increasing of the surface magnetic field, it becomes 
more uniform and the level of activity changes in this case in less degrees 
than on stars with strong local fields concentrated in large spots.
}

\begin{document}

\flushbottom % Makes all text pages the same height

\maketitle % Print the title and abstract box

\tableofcontents % Print the contents section

\thispagestyle{empty} % Removes page numbering from the first page

\section{INTRODUCTION}

Study the activity of main sequence stars is fundamental 
importance of both for understanding the
development and scales of active phenomena, nature
of their generation, as well as some the related to stellar
activity issues concerning to studying planets, the origin of life 
(a habitable zone), large-scale cataclysms,
an existence of civilizations.

At present, the instrumental capabilities make it
possible to determine the activity indices of G--K stars
in all layers of the atmosphere and to measure their
magnetic fields. Dynamo processes occurring in the
convection zone lead to an increase of the magnetic
field and its exit to the surface and in upper layers of
the atmosphere.

The relationship of parameters of active regions
with the physical parameters of stars and rotation has
been well studied for solar-type stars with periods of
more than 5 days (Baliunas et al., 1995; Ol\'ah et al.,
2016; Distefano et al., 2017). On fast-rotating stars
with magnetic fields of about 4 kG and more, the
activity is less studied, but it has been established that
for such strong magnetic fields the relationship
between rotation and activity level is disrupt, and an
increasing of the rotation does not lead to increase
activity indices, i.e., a saturation level is reached
(Reiners et al., 2009; Shulyak et al., 2017; 2019). 
Photometric data for fast-rotating stars were obtained 
for limited time intervals and for a small set of stars. 
Currently, in connection with the search for exoplanets,
the photometric behavior of M dwarfs is being actively
studied, but in most cases it is data for 7--15 years
(Savanov, 2012; Su\'arez Mascare\~no et al., 2016; 
Distefano et al., 2017; Iba\~nez Bustos et al., 2019). 
The convective zone of M-dwarfs is deeper than in G--K
dwarfs, and dwarfs of the spectral class M3.5 and later
become fully convective (Chabrier \& Baraffe, 1997).
Thus, the conditions for the development of dynamos
and magnetic fields in main-sequence stars are different, 
and to study the activity of stars with different
parameters, the accumulation of data on activity indices 
and magnetic fields remains relevant.

In this paper, we present the results of the search
for cycles at four stars -- OU Gem (dK2), EQ Vir
(dK5), V1005 Ori (dK0), AU Mic (dM1). These are stars
with fast rotation and magnetic fields of 2--2.6~kG
(Shulyak et al., 2017). For them, the several decade's
photometric series were first obtained and studied,
and possible activity cycles were revealed. The nature
of the relationship between the parameters of the cycle
and the average magnetic field for fast-rotating stars
(young age) that have not reached saturation is considered 
by the example of the named group of stars,
supplemented by several previously studied objects.

\section[LONG-TERM LIGHT CURVES AND \\ PHOTOMETRIC DATA 
SOURCES]{LONG-TERM LIGHT CURVES AND PHOTOMETRIC DATA SOURCES}

The activity cycle associated with the development
of sunspots lasts about 11 years. As observations have
shown, for solar-type stars and smaller masses, longer
cycles are possible, and therefore, to search for them it
is necessary to study photometric series at intervals of
more than 10 years. For the selected stars long-term
light curves were formed according to data obtained in
photographic archives, published results of photoelectric
photometry and taken from the
\href{http://cds.u-strasbg.fr/cgi-bin/Dic-Simbad?HIP}{Hipparcos databases},
ASAS (Pojmanski, 1997) and KWS 
(\href{http://kws.cetus-net.org/~maehara/VSdata.py}{Kamogata/Kiso/Kyoto Wide-field Survey}). 
In the course
of statistical processing of arrays of $V$-values selected
from the indicated databases, erroneous values and
flares were identified and removed, average brightness
values for each observation date, average annual values
and errors of their determination were determined. The
mean square errors of $V_{yr}$ -- values did not exceed 0.07$^m$
according to photographic data, 0.015$^m$ according to
photoelectric data, 0.015$^m$, 0.03$^m$, 0.028$^m$ according to
Hipparcos, ASAS and KWS, respectively.

\begin{figure*}[!th] %%% Figure 1
\centering
%%%\vspace*{-0.9cm}
\includegraphics[width=\linewidth]{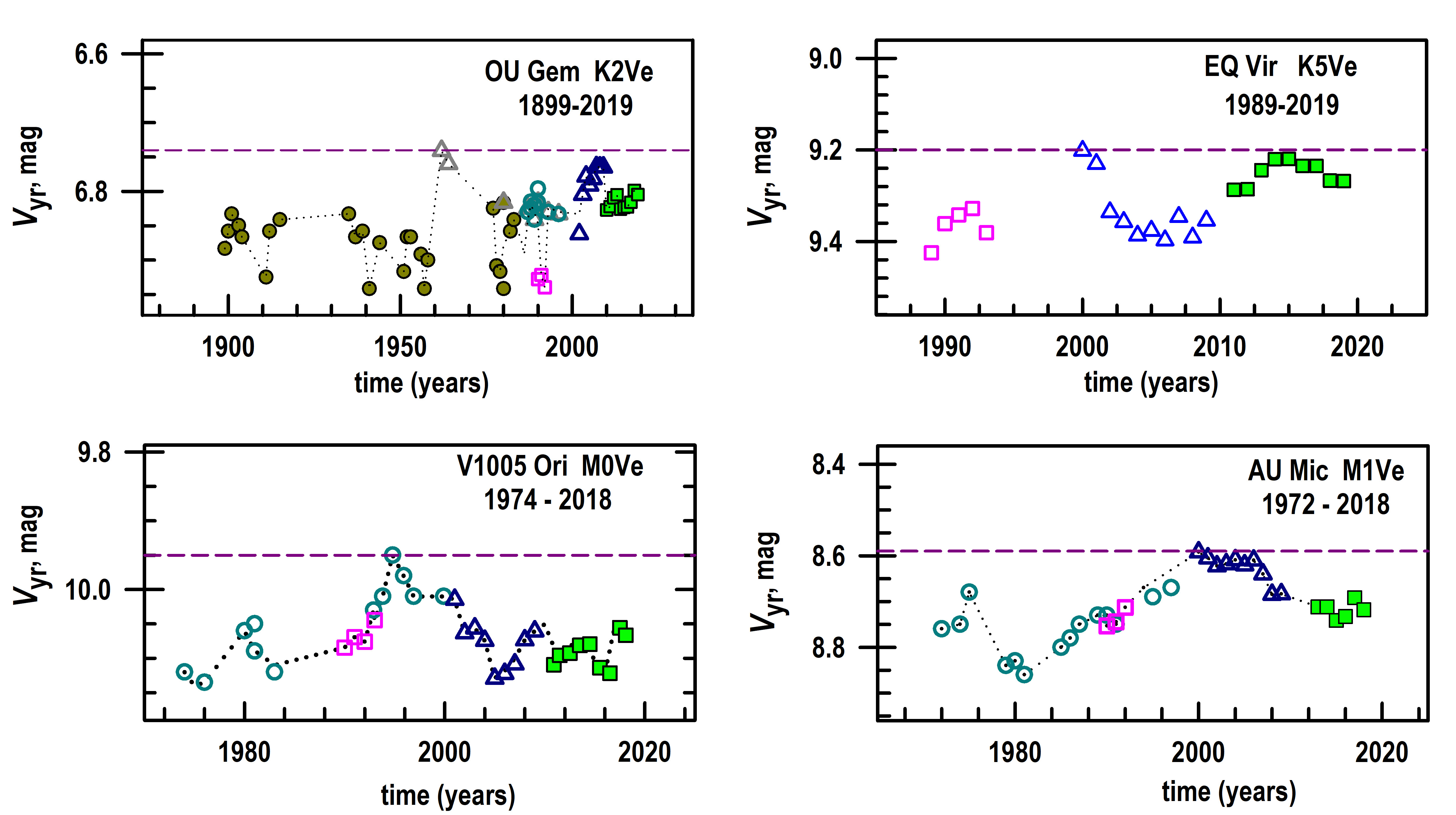}
%%%\vspace*{0.2cm}
\caption{
Changes in the average annual brightness at long-term intervals for active red dwarfs 
OU Gem, EQ Vir, V1005 Ori, AU Mic.
Different symbols indicate the following data sources: filled circles -- photographic collection of Sternberg Astronomical Institute, Moscow State University (Bondar', 1995), open circles -- published photometry data by Cutispoto (1995,1998) and Cutispoto et al. (2001, 2003), open squares -- catalog of Hipparcos satellite data, open triangles -- ASAS photometric catalog, filled squares -- KWS database. The dashed line marks the level of maximum brightness on the presented observational 
interval.
}
\label{Figure1}
%\end{center}
\end{figure*}

For the star OU Gem it were used photographic
magnitudes for 1899--1988 (Bondar', 1995), electrophotometry
collected by Alekseev (2001) obtaining
from his observations at the 1.25~m AZT-11 telescope
and taken from publications (1972--1988), and results
from databases Hipparcos (1990--1992), ASAS
(2000--2009) and KWS (2010--2019). The $B$-magnitudes
added from photographic data were converted to
$V$-values according to the corresponded $B-V$-values.

The complete light curve in Fig.~1 covers the time
interval of 120 years, from 1899 to 2019, the number of
average annual $V$-values $N = 56$. The brightness
changes from year to year amounted to 0.2$^m$; the maximum
level (marked by a dashed line) was reached in
1962 and 1964 (Argue, 1963) and 2004 (ASAS).

The behavior of the average annual brightness of
EQ Vir is presented on the interval 1989--2019 ($N =
25$) by photometry from the Hipparcos (1989--1993),
ASAS (2000--2009) and KWS databases (2010--2019)
(Fig.~1). Changes in the brightness amounted of 0.22$^m$.
The photographic data for 1914--1988 (Bondar', 1995)
are not numerous and very sparsed, they were not
included in the data array for a search for the cycle, but
we took into account, that on this interval and also in
1993--1998 (Alekseev, 2001), the brightness did not
reach its maximum of $V = 9.20^m$ that was found from
ASAS and KWS data.

For star V1005 Ori, the results of 1974--1976
obtained by Bopp and Espenak (1977), Bopp et al.
(1978), data for 1980--1981 by Byrne et al. (1984) and
1992--1999 (Alekseev, 2001) were taken, and $V$-values
from Hipparcos (1990--1993), ASAS (2001--2009)
and KWS (2010--2019) as well. The average annual
values from the Hipparcos catalog were recalculated,
$V_{yr} = V_{Hip} - 0.11$ correspondently to the data from
(Alekseev, 2001). The light curve in Fig.~1 includes 34
epochs of observations. Changes in the average annual
brightness from 1974 to 2019 amounted to 0.19$^m$, on
the entire studied interval the maximum brightness
level of 9.25$^m$ was observed only once, in 1994.

The AU Mic light curve (Fig.~1) covers the interval
from 1972 to 2018. The yearly mean values ($N = 32$)
were obtained from data by Cutispoto (1995; 1998),
Cutispoto et al. (2001; 2003) Hipparcos (1990--1992),
ASAS (2001--2009) and KWS (2013--2018). Changes
in the average annual brightness were 0.27$^m$, and the
maximum brightness was 8.59$^m$ in 2000. According to
archival data from Phillips and Hartmann (1978), this
level was not reached on the span from 1913 to 1971. In
Table~1, we present the maximum magnitudes of stars
obtained by us in the studied time interval and the data
from the SIMBAD.

\begin{table*}[!th] %%% Table 1
\centering
%%%\vspace*{-0.9cm}
%%%\vspace*{0.2cm}
\caption{
Maximal $V$-magnitudes for investigated stars 
}
\par\medskip
\begin{tabular}{|c|c|c|c|c|}
\hline
star name	& time      & $V$, mag   & $V_{max}$, mag & years of maximum \\
            & interval	& (SIMBAD)   &                 & in $V_{yr}$-values \\
\hline
OU Gem	& 1899--2019	& ---	& 6.74 & 1962, 2007 \\
\hline
EQ Vir	& 1989--2019	& 9.37	& 9.20	& 2000, 2015 \\
\hline
V1005 Ori &	1974--2018 &	10.11	& 9.95	& 1994 \\
\hline
AU Mic &	1972--2018 &	8.63	& 8.59	& 2000 \\
\hline
\end{tabular}

\label{Table1}
%\end{center}
\end{table*}

\section[SEARCH FOR POSSIBLE CYCLES IN \\
PHOTOSPHERES OF THE SELECTED STARS]{SEARCH 
FOR POSSIBLE CYCLES IN PHOTOSPHERES OF THE SELECTED STARS}

Long-term light curves make it possible to determine
the epochs of maximum brightness and the
amplitude of variability. Changes in the average
annual brightness for active solar-type stars and red
dwarfs are explained in the frameworks of the concept
of the development of surface starspots. The epochs of
minimum brightness correspond to the maximum
spottedness (maximum activity), and spots on the surface
of the star are absent or their number is minimal
in the years of the greatest brightness. The interval
between successive epochs of the star's maximum
brightness (activity minima) shows the duration of the
possible cycle of its photospheric activity. The stellar
activity cycle can last several decades and, according
to available observations, it is rarely possible to trace
the development of several cycles. Therefore, the
found long-term cycles should now be considered
possible or suspected, but their search allows us to
identify stars with cycles and irregular activity, which is
important for studying the nature of the magnetic
activity of stars.

The search for periods in brightness variations for the stars considered 
by us was performed  used the methods of Yurkevich and Hartley as described 
early by Bondar' (2019).
For OU Gem, 
for a number of average annual $V$-values, the presence of periods
of less than 60 years was considered; the best data convolution
was obtained for a period of 41.6 years. Figure~3
shows the phase curve for this period and the approximation
of the average annual values in bins of 0.1 phases by
a 5$^{\rm th}$ degree polynomial (dashed line), the initial epoch
is chosen arbitrarily, $T_0$ = 1899. The amplitude of the
possible cycle is 0.09$^m$.

\begin{figure*}[!th] %%% Figure 2
\centering
%%%\vspace*{-0.9cm}
\includegraphics[width=\linewidth]{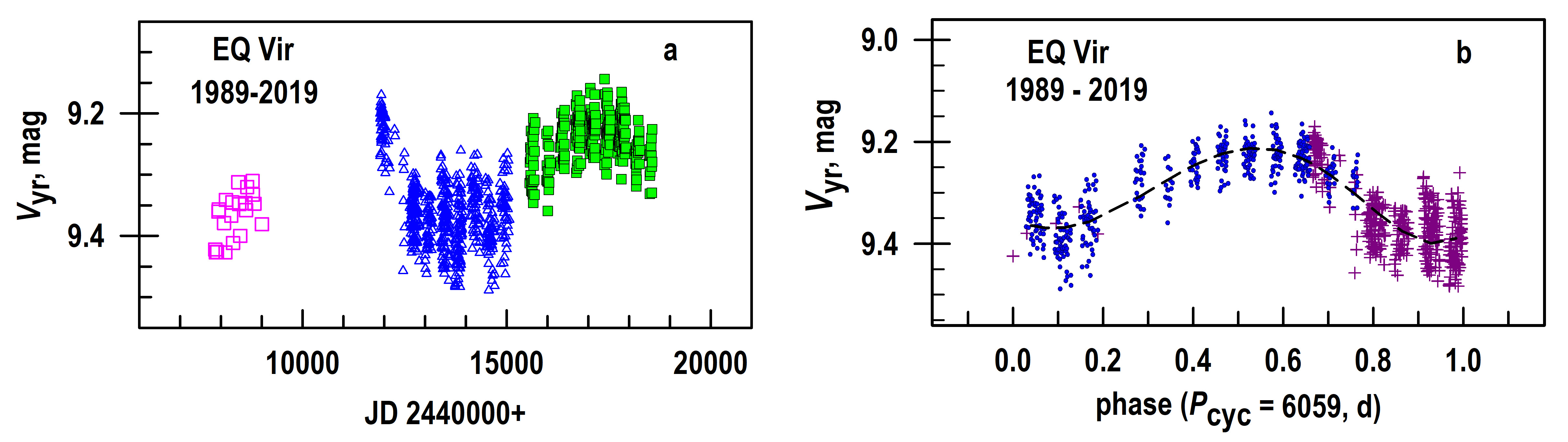}
%%%\vspace*{0.2cm}
\caption{
Activity cycle in photosphere of EQ Vir  from $V$-data obtained  in 1989--2019.
a) Average $V$-values for the date of observations according to the data from  Hipparcos, ASAS and KWS, description to symbols as in Fig.~1.
b) The convolution of the averaged $V$-data with $T_0 (JD) = 2447868$ and the found cycle of 6059.53 d shows two consecutive cycles marked with dots and crosses.
}
\label{Figure2}
%\end{center}
\end{figure*}

For EQ Vir, we analyzed a series of averaged $V$-magnitudes
for each date of observations (Fig.~2).
According to the ASAS and KWS data, the possible
cycle ($P_{cyc}$) on the interval JD2451900--2458569 was
determined as equals to 6095.62 days. The Hipparcos
data are corresponded to the convolution with this
period, and we have precised the value of the possible
cycle for the entire data on the interval 
JD 2447868--2458569 ($T_0 = 2447868.49$, $N = 874$). With the obtained
$P_{cyc} = 6059.53\;$days (16.59 years), two waves of this cycle
were observed. They are shown by the corresponding
icons in the left graph of Fig.~2. It can be seen that the
shape and phase of the cycle are not changed. A convolution
of the mean yearly $V$-values is shown in Fig.~3,
$T_0 = 1989$, the approximation by bins equal to 0.1 of the
phase represents the shape of the found cycle and its
amplitude $A_{cyc} = 0.19^m$.

\begin{figure*}[!th] %%% Figure 3
\centering
%%%\vspace*{-0.9cm}
\includegraphics[width=\linewidth]{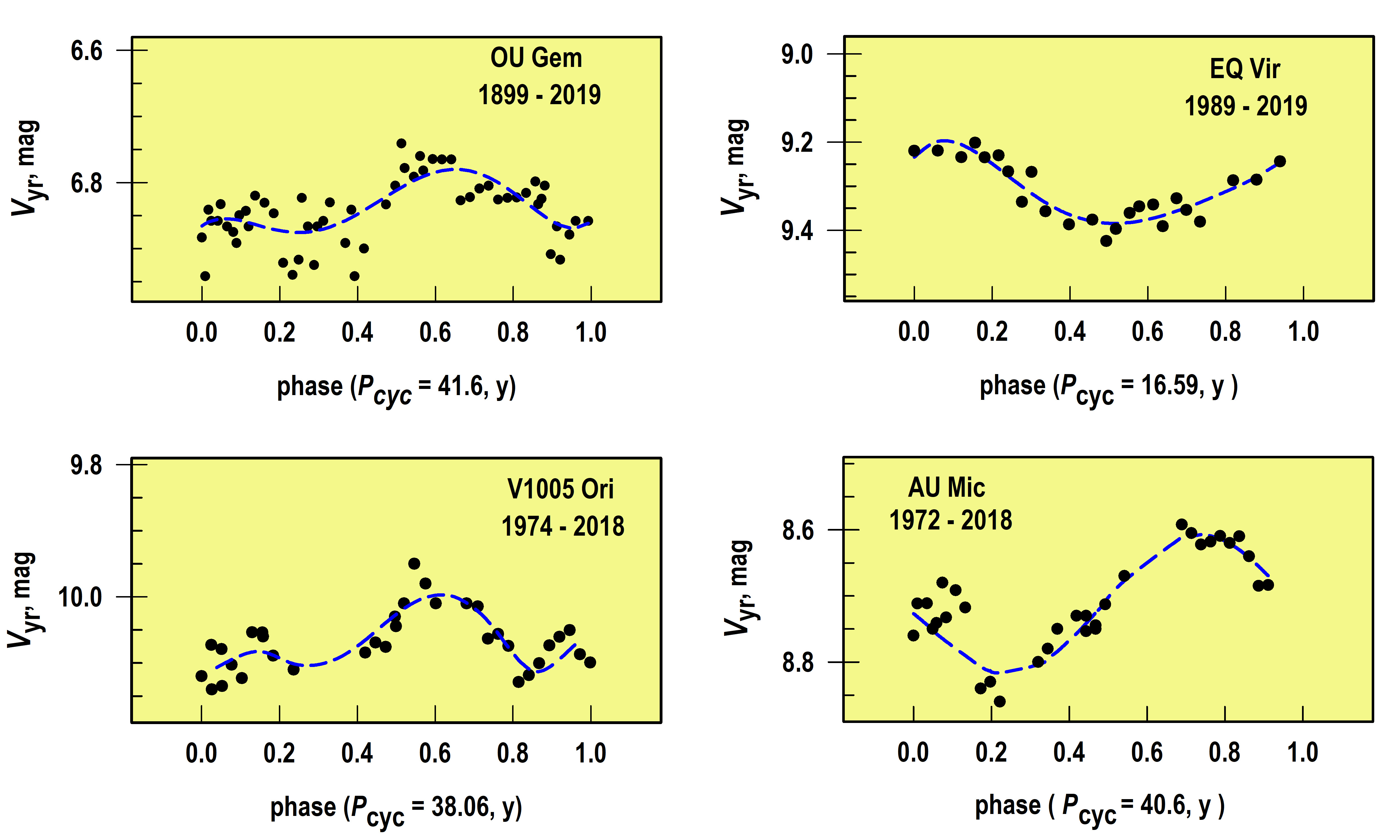}
%%%\vspace*{0.2cm}
\caption{
Long-term activity cycles in the studied stars.
Filled circles present average annual $V$-values, dashed lines draw an approximation 
of phase curves by a high order polynomial and show the shape and amplitude of a possible cycle.
}
\label{Figure3}
%\end{center}
\end{figure*}

For the M-dwarfs V1005 Ori and AU Mic, the
search for possible periods is complicated by the fact
that only one epoch of maximum brightness was
recorded on the studied intervals for these stars. Periodogram
analysis was performed for a series of average
annual $V$-values and for the row obtained by approximating
the light curve with a high degree polynomial.
The best convolution of the yearly mean $V$-values was
obtained for V1005 Ori with the period of 38.06 years
($T_0 = 1974$), for AU Mic -- with a period of 40.6 years
($T_0 = 1972$) (Fig. 3). The amplitude of the suspected
cycles is 0.13$^m$ and 0.21$^m$, respectively. Approximation
by polynomials has been carried out according to the
average values in bins equal to 0.1 of the phase of the
accepted periods. The parameters of the discussed
activity cycles are shown in Table~2.

\begin{table*}[!th] %%% Table 2
\centering
%%%\vspace*{-0.9cm}
%%%\vspace*{0.2cm}
\caption{
Parameters of photospheric activity cycles and magnetic fields for the researched red dwarfs
}
\par\medskip
\begin{tabular}{|c|c|c|c|c|c|c|c|}
\hline
& & & & & & & \\
star name	& spectral type & $B-V$	& $P_{cyc}, y $ & $A_{cyc}$, mag & ref & $\langle B \rangle$, G & ref\\
& & & & & & & \\
\hline
LQ Hya	& K2.0	& 0.91	& 18 & 0.27	& 6	& 2450	& 1\\
\hline
OU Gem	 & K2.0	 & 0.95	 & 41.6	 & 0.09	 & 9	 & 2400	& 2\\
\hline
EQ Vir	& K5.0	& 1.18	& 16.6	& 0.19	& 9	& 2000	& 3\\
\hline
V833 Tau & K5.0	 & 1.04	 & 78	 & 0.6	 & 7	 & 1400	 & 1\\
\hline
V1005Ori & M0.0	 & 1.37	 & 38.1	 & 0.13	 & 9	 & 2600	 & 4\\
\hline
AU Mic & M1.0	 & 1.32	 & 40.6	 & 0.21	 & 9	 & 2300	 & 5\\
\hline
DT Vir & M2.0	 & 1.47	 & 31.45 & 0.15	 & 8	 & 2600	 & 4\\
\hline
AD Leo	 & M3.5	 & 1.55	 & 27	 & 0.16	 & 6	 & 3100	 & 4\\
\hline
DX Cnc	& M6.5	& 2.05	 & 2.67	 & 0.10	 & 8	 & 3200	 & 4 \\
\hline
\end{tabular}
\par\bigskip
References:  (1) Saar S. H., 1996; (2) Saar \& Linsky, 1986;  (3) Saar S. H. et al. 1986;  (4) Shulyak et al., 2017; (5) Saar S. H., 1994; (6) Alekseev \& Kozhevnikova 2017; (7) Bondar' 2015; (8) Bondar'  2019; (9) this paper 

\label{Table2}
\end{table*}

\section{ACTIVITY CYCLES AND MAGNETIC FIELDS}

The spots on star's surface are regions of an
enhanced magnetic field, and therefore, the surface
magnetic field of active stars is inhomogeneous. In
this case, the average magnetic field is calculated as
the sum of the weighted values of the magnetic field
strength $\langle B\rangle = \Sigma|B_i | f_i$ 
in each of its elements with the
coefficient $f_i$, the filling-factor, which expresses the
area of the region covered by the magnetic field with
the intensity $|B|_i$ in fractions of stellar surface area
(Shulyak et al., 2014; 2017). In these works, the
authors give a method for determining the average surface
magnetic field on low-mass stars.

The G--M dwarfs studied by us were selected from
the list of stars with measured magnetic fields
(Shulyak et al., 2017). According to the data in the list,
the average magnetic field is amplified in rapidly
rotating fully convective M stars, for some of them 
$\langle B\rangle$
exceeds 4~kG. We tried to consider the effect of a surface
magnetic field on the development of cyclic photospheric
activity according to photometric observations.
As is known, the amplitude of the cycle is related
to the area occupied by the spots: 
$A^2_{cyc} \sim S_{spot}$. 
To consider the relationship between cycle parameters and
$\langle B\rangle$, we added in our sample 5 K--M dwarfs, 
parameters of their cycles were taken from (Bondar', 2019).
The list of stars, parameters of cycles and the mean
magnetic fields are presented in Table~2.

\begin{figure}[!th] %%% Figure 4
\centering
%%%\vspace*{-0.9cm}
\includegraphics[width=\linewidth]{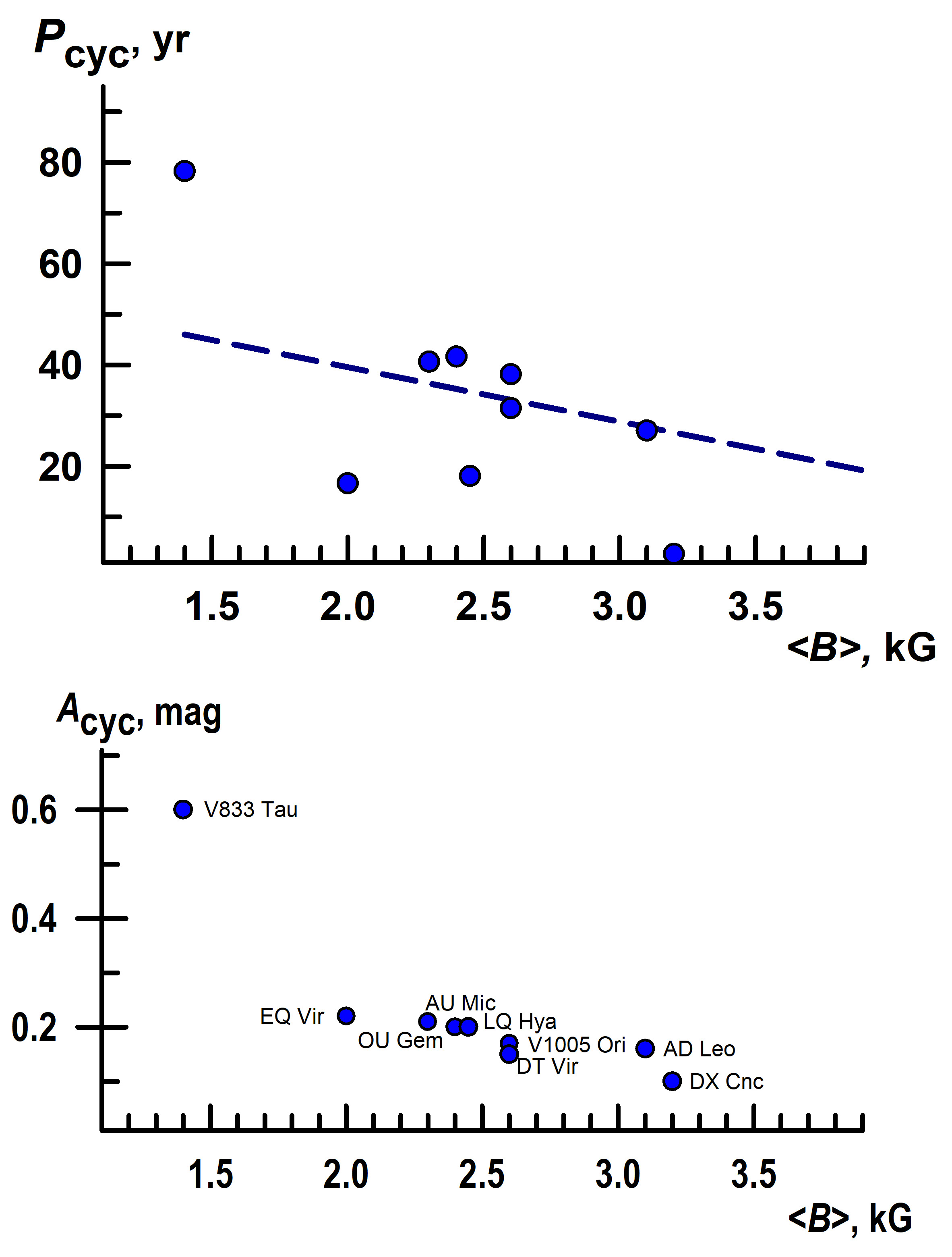}
%%%\vspace*{0.2cm}
\caption{
The relationship between the surface magnetic field and the activity cycle parameters of fast-rotating red dwarfs.
}
\label{Figure4}
%\end{center}
\end{figure}

Besides V833 Tau, the averaged surface magnetic
fields of the considered stars are 2--4~kG. For stars
with such fields, the amplitude and duration of the
cycle are increased with a decreasing of 
$\langle B\rangle$ that is displayed
on Fig.~4. The limited number of the considered
stars allows us only to propose the observed tendency.
The deviations from the regression line that are
noticeable in the graphs, as well as deviations of the
cycle parameters for V833 Tau, can be caused both by
errors in their determination and the actual existence
of groups of stars with a specific type of relationship
between the magnetic field and parameters of cycles.
The question of whether it is possible to expect the
development of long and high-amplitude cycles on
stars with surface magnetic fields $<2\;$kG requires further
study.

\section{CONCLUSION}

The behavior of the average annual brightness of two
K-dwarfs OU Gem and EQ Vir and the early M-dwarfs
V1005~Ori and AU~Mic was first studied on time span
covers of several decades using photometry from databases
for 1989--2019. Long-term series allowed us to
determine maximum brightness values for the studied
stars, all of them differ from those given in the astronomical
SIMBAD database. The variability of the
average annual brightness of all stars is cyclical.

The durations of possible cycles for OU~Gem,
V1005~Ori and AU~Mic are 40--42 years, for EQ~Vir --
16.6 years, and amplitudes are not more than 0.2$^m$.
Measurements of the magnetic fields of the stars studied
by us showed that the average surface magnetic field
strengths are several kilogauss. We exami\-ned the nature
of the relationship between the parameters of the cycle,
its duration and amplitude, and the value of $\langle B\rangle$, 
including another 5 stars with strong magnetic fields.

A noticeable tendency is an increase of the amplitude
and duration of a cycle with a decrease in the
value of $\langle B\rangle$. This may be due to the fact that 
with an increase in the surface magnetic field, it becomes
more uniform and the level of activity changes to a
lesser extent than in stars with strong local fields concentrated
in large spots. It is known that the surface of
M-dwarfs is covered with small but numerous spots.

\section*{ACKNOWLEDGMENTS}

The authors are grateful to A.A.~Shlyapnikov for help in
working with databases, also to Z.A.~Taloverova for technical
help in preparation of text.
In the work we used information from the SIMBAD
database, photometric data from the Hipparcos, the International
Variable Star Index (VSX) database, the 
Kamogata-Kiso-Kyoto Wide-Field Survey. The authors are thankful
to all the staff providing the replenishment of these databases
and access to them, as well as the developers of the
AVE software package.

\section*{FUNDING}
This work was carried out with the partial support of the
Russian Foundation for Basic Research (grant 19-02-
00191a and 18-52-06002~Az\_a).

\section*{CONFLICT OF INTEREST}
The authors declare that they have no conflicts of interest.

%%%\phantomsection
\bibliographystyle{unsrt}
%\bibliography{sample}

\end{document}